# K. Designing for interpersonal museum experiences
**Anders Sundnes Løvlie, Lina Eklund, Annika Waern, Karin Ryding & Paulina Rajkowska**

*Dr Anders Sundnes Løvlie is Associate Professor at the IT University of Copenhagen and works with experience design, play and media. He was coordinator for the Gift project (gifting.digital).*
*Dr Lina Eklund is Lector in Human Computer Interaction at Uppsala University, Sweden and works on digital sociality, with a particular focus on museums and games.*
*Dr Annika Waern is Professor in Human Computer Interaction at Uppsala University, Sweden and works on design for embodied play and playfulness over a number of domains.*
*Karin Ryding is a PhD fellow at the IT University of Copenhagen and works on play design with a particular focus on performativity, relationality and affect.*
*Paulina Rajkowska is a PhD student at Uppsala University. She is a lecturer teaching within the fields of HCI, Social Media and Communication Studies.*

## Introduction

What does the age of participation look like from the perspective of a museum visitor? Arguably, the concept of participative experiences is already so deeply ingrained in our culture that we may not even think about it as participation. Museum visitors engage in a number of activities, of which observing the exhibits is only one part. Since most visitors come to the museum together with someone else, they spend time and attention on the people they came with, and often the needs of the group are given priority over individual preferences (McManus 1989). Thus visitors will be socialising with other visitors in their group, looking after their children, visiting the museum café, shopping in the gift shop, etc. They are also engaged in things that lie outside the museum's offer, such as googling for information, exchanging messages and updates with their peers, taking selfies and posting them on social media, playing games, and so on. According to Blud, "interaction between visitors may be as important as interaction between the visitor and the exhibit" (Blud 1990, p.43). The museum visit is (always already) a highly participative experience, with or without the museum's blessing and not necessarily to its benefit. How can museums tap into these activities - and make themselves relevant to visitors?

The first thing to notice is that participation, from this perspective, does not necessarily take the form of a "dialogue" between the museum and its visitors. Rather, the dialogue takes place between visitors who are interacting with each other. The museum is merely the context, the pretext or perhaps the backdrop for that interaction. While this may sound like a pessimistic perspective for museum curators, in this chapter we will try to approach this constructively, as a design opportunity. Could it be productive for the museum to consider itself not only as a disseminator of knowledge, but also as the facilitator of participative activities between visitors?

Graham Black (2021) *Museums and the Challenge of Change: old institutions in a new world*
Section 3
    K. Løvlie et al: Designing for Interpersonal Museum Experiences
Submission 01/08/2020

In what follows, we will outline a range of practical design projects that serve as examples of this approach. These projects were part of the European Union funded Horizon2020 project GIFT, a cross-disciplinary collaboration between researchers, artists, designers and many international museums and heritage organisations, exploring the concept of interpersonal museum experiences (see https://gifting.digital/). What the projects have in common is that they build on visitors co-creating and sharing their own narratives in the museum context. We suggest that these projects demonstrate a spectrum of possibilities: From experiences that take place almost without any museum involvement, to those that give museums a role in curating these narratives.

First, let us have a closer look at how museum visitors interact with each other.

## Interpersonal Experiences with/in the Museum

Museums are social spaces. Visitors go there with friends and family for a day out together (McManus 1989). As a social leisure activity, reaffirming social relationships is an underlying motivator and this has a significant impact on any visit. Visitors often operate under a form of social contract, a responsibility to their companions to maintain the social framing of the visit. In a recent study we followed groups of friends from Generation Z (in this case, predominantly university students), as they visited Gustavianum, a University history museum located in Uppsala, Sweden (Eklund in press). We looked at how social meaning-making occurred as visitors oriented themselves towards the exhibits, creating new meaning of relevance to them as friends. For example, two notebooks by former students from the 1600s and the 1800s were remarked upon by many visitor groups. They marvelled at the masses of notes and reflected on their own note-taking practices on digital devices. Through reflecting on the academic practices of old, the visitors were able to reaffirm being students together.

Museum visits can be an opportunity to strengthen relationships. During a visit people engage in many different types of sub-activities, seamlessly moving back and forth between social interaction, play, exploration, navigation, reading, and so on while affirming their friendships. Through social recontextualization visitors reinterpret and frame museum artefacts to become relevant to them. Visitors will reference shared histories, memories, and identities such as discussing a student party while looking at a Viking drinking bowl or engaging in the sharing of knowledge between each other rather than reading museum texts. Visitors may go as far as to recontextualize through role play; for example, by enacting the famous prow scene from Titanic in an old Viking ship.

Exhibited cultural heritage might be defined by the museum (Calcagno & Biscaro, 2012), yet the audience in turn add new meaning, bringing with them their lived experiences. Through the mundane and every-day, the ancient and thus 'foreign' objects are untangled and made sense of; are recontextualized. Humour and play are key mechanisms in this process, making jokes, role-playing with and through objects and locations supports sociability and engagement with the visit. Recognizing what people do in museums is the first step to creating participative museum experiences.

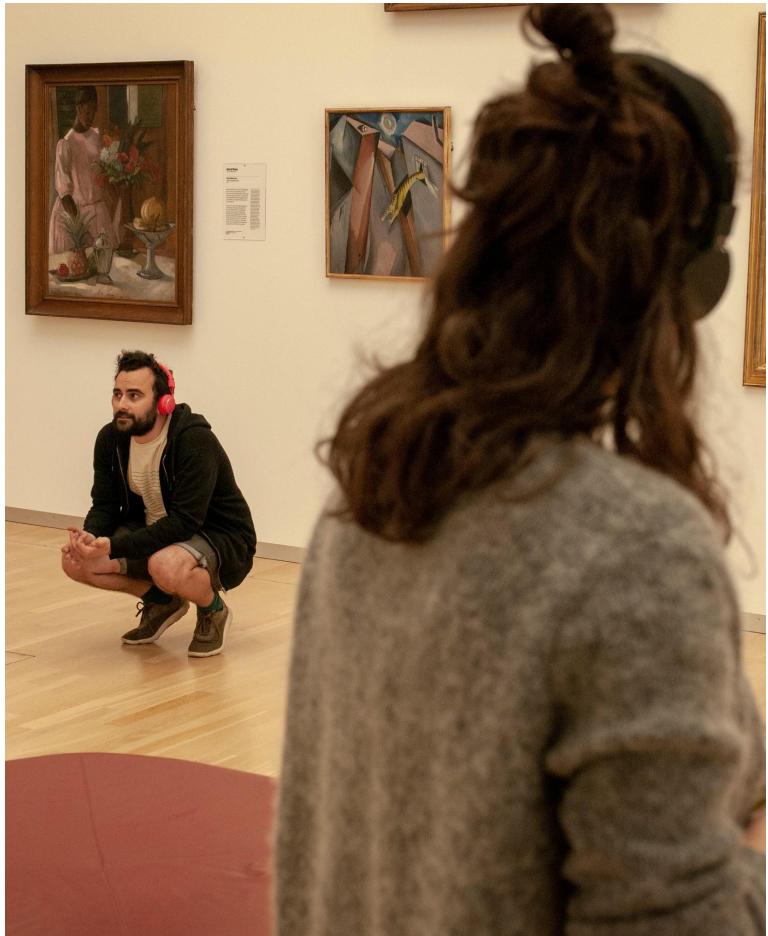

Figure K.1: Participants playing *Never Let Me Go* in the National Gallery of Denmark, Copenhagen. Photo: Johan Peter Jønsson.

## Designing for the Interpersonal: Play and Gifting

> It was like the place came alive a bit more to me. I especially remember one of the first things you said was: 'Imagine if this is looking back at you'. I felt like all the pictures were staring at me.

*Never Let Me Go* participant, cited in Ryding & Fritsch (in press)

Next, we turn to two designs that facilitate interpersonal experiences in the museum - and through that, help visitors see the museum in a new light. Museums often wish to tap into play and games, in order to facilitate new modes of visitor engagement. Play is often a social activity - as are museum visits. However, museums also wish for visitors to spend time exploring and learning about exhibits - often idealised as a contemplative and "transformative" experience (Soren 2009). Can these contrasting ideals be combined? The design experiment *Never Let Me Go* (Ryding 2020; Ryding & Fritsch in press) presents an interesting approach to facilitating experiences that are simultaneously social and introspective. *Never Let Me Go* is a mobile game for two players, for use in art galleries. One player acts as "controller", taking command of the other player who acts as "avatar". The controller can send a range of commands to the other player, who will hear them in headphones. The commands range from "Basic commands" such as "Explore", "Follow", and "Wait", to more ambiguous suggestions such as "Become part of this" or "Imagine that this is looking back at you". Some commands are questions: "Can you feel the longing in this?"

The application was tested, with permission, in the National Gallery of Denmark. Trial visitors described it as an intimate experience which sets the interpersonal relationship between the two players up as a lens through which they may achieve a highly personal experience of the museum. Through this, the game also offers the visitors a new way to look at art: "It felt stimulating. A way of asking new questions" (Ryding & Fritsch in press). *Never Let Me Go* was created without any involvement from the museum, as an independent, reflective design. The app interface is also completely agnostic of the specific museum, so the game can in principle be played in any kind of museum - although it has been designed with art museums and galleries in mind.

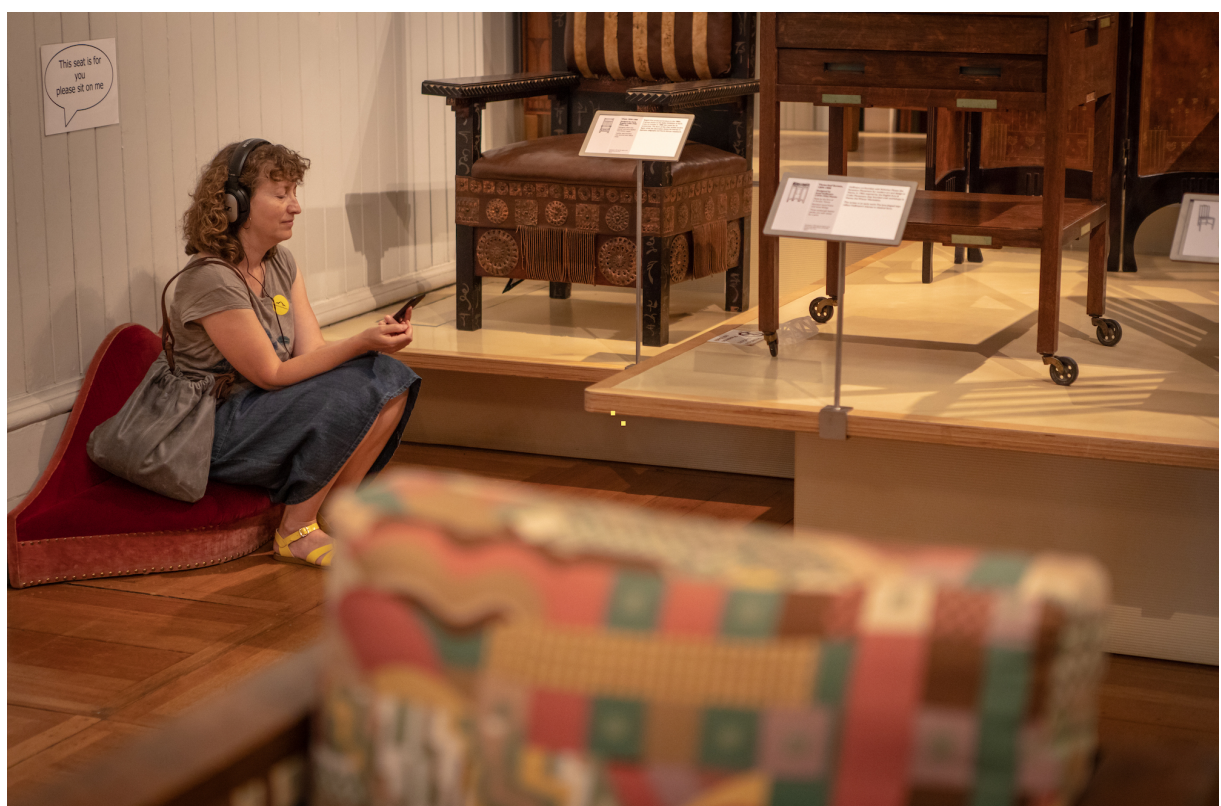

Figure K.2: Participant using the *Gift* app by Blast Theory in Brighton Museum. Photo: Charlie Johnson.

The interpersonal relation between museum visitors is also the primary focus of the British artists group Blast Theory's app *Gift*, which greets museum visitors with the words: "You are going to make a gift for someone special. They might be next to you right now, they might be on the other side of the world." The app asks museum visitors to see the museum through the eyes of their 'someone special' and select three objects that they think that person would like to put in a "gift". For each object, they may include a photo of the object and record a personal audio message. The recipient may open the gift at home as a purely digital experience, or come to the museum and experience the gift as a small, personalised "tour". The artists compared the app to the practice of making mixtapes with music: "Ever made someone a mixtape? How about with objects from a museum?"

Similarly to *Never Let Me Go*, the users of *Gift* have described the experience as highly personal, offering them a meaningful way to engage with their personal relationship to a special other; but also simultaneously as giving them a fresh perspective on the museum and its exhibits, seeing them with "new eyes" (Spence et al. 2019). Consider, for example, the gift that the teenager "Kristin" made for her mother, a picture of the painting "Alice in Wonderland" by George Dunlop Leslie, along with an audio message explaining why she thought her mother might be interested in this painting:

> "So, this picture is called Alice in Wonderland, from 1879, and the sofa reminded me a lot of grandma's sofa with the dolls. And the poem says that this is a big sister reading to her little sister, and I think you can imagine me and Leni sitting like this and her reading to me my favourite story."

<div align="right">Løvlie et al., 2019</div>

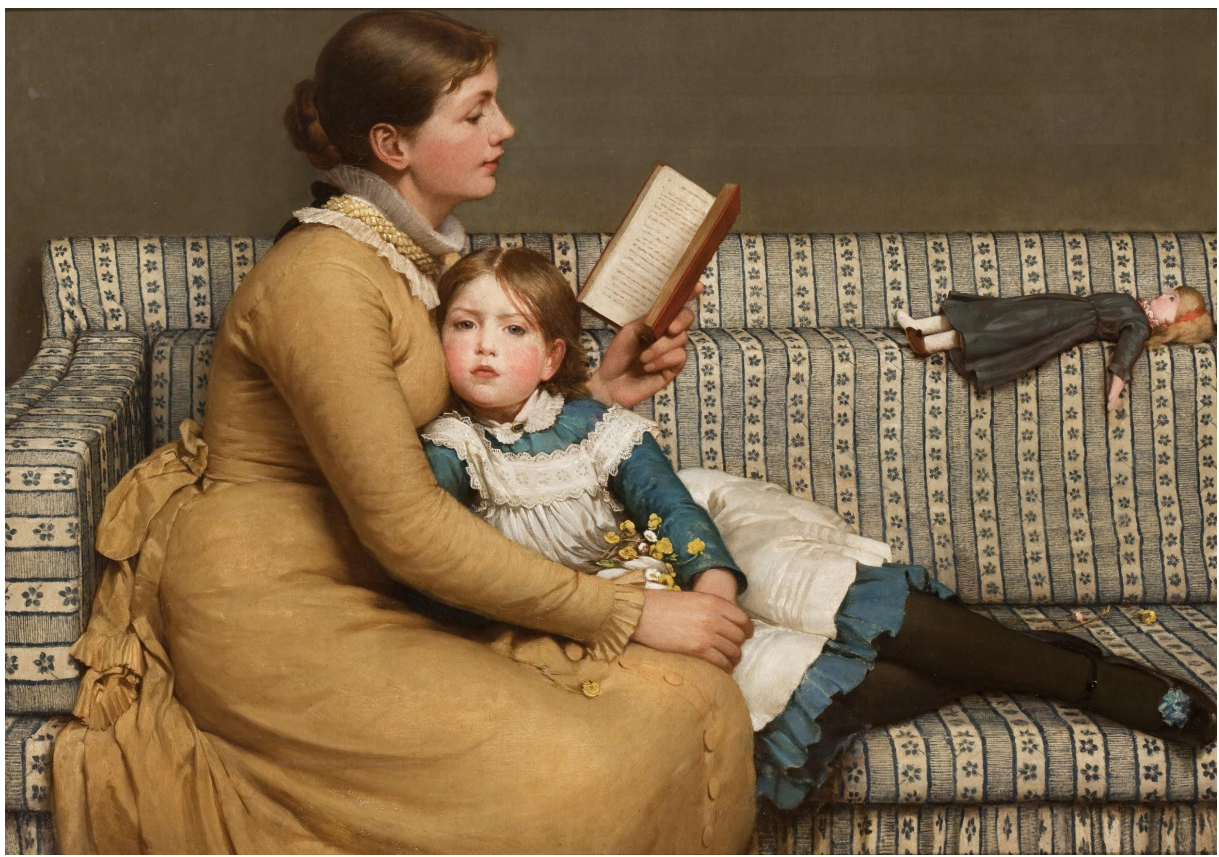

Figure K.3: *Alice in Wonderland*. Oil painting by George Dunlop Leslie, c1879. Royal Pavilion & Museums, Brighton & Hove (Creative Commons CC-BY-SA).

While the gift is a personal message from the giver to the recipient and as such their personal relation is at the heart of the exchange, it also offers them a lens through which to see the museum object - and engage with it on their own terms.

*Gift* was created in collaboration with Brighton Museum and has later been commissioned by the Munch Museum in Oslo, Norway, and the Museum of Applied Art in Belgrade, Serbia. The app was created as an independent art piece and is, just as *Never Let Me Go*, mostly museum-agnostic. However, Blast Theory offers to work on commission to tailor it to specific museums, which allows a slightly more museum-centric perspective than *Never Let Me Go*.

## Museum-Curated Interpersonal Experiences

How about the more traditional type of interaction that museum visitors have with museum objects - looking at the objects and reading about them - can this interaction be turned into an interpersonal experience? Image recognition technology has greatly expanded the possibilities for creating mobile guide applications for museums. Apps such as Smartify, Vizgu and Magnus invite visitors to simply point their smartphone camera at an artwork in the museum, and the app will tell them what they are looking at. However, while it is a trivial task for an algorithm to collect metadata about an artwork from the museum's database, it is far from trivial to turn this data into an engaging experience for the visitor. This challenge is made greater by the (perhaps disappointing) fact that museum visitors spend surprisingly short time studying canonical works of art - typically around 15-30 seconds (Smith et al. 2017).

The app *One Minute Experience* addresses this challenge by presenting the visitor interpretive text about an artwork in the form of a short story that can be read in one minute or less. The app comes with a story editor which helps curators write highly condensed narratives around a series of prompts. But not just curators: Why not also let other voices in? There are communities with interesting stories to tell. What perspective could art students at a local university offer? Or how about folks from the LGBT+ community, ethnic minorities, immigrants, war veterans - or children?

Working with Brighton Museum, in November 2019 we invited interested individuals from the museum's networks to contribute their stories to the app. Through three days of workshops we gathered 24 stories, about a variety of objects from queer fashion to archaeological artefacts and art paintings (see example in Figure K.4a - f).

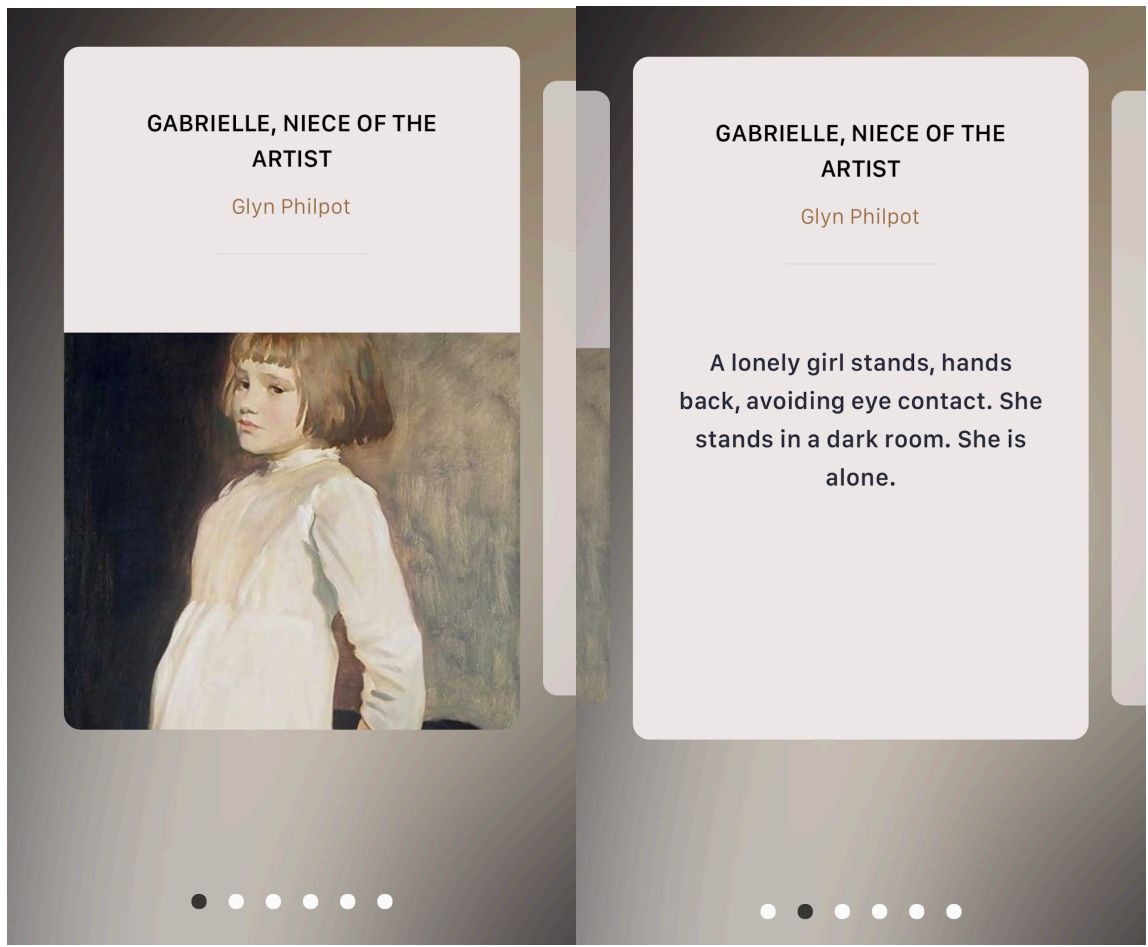

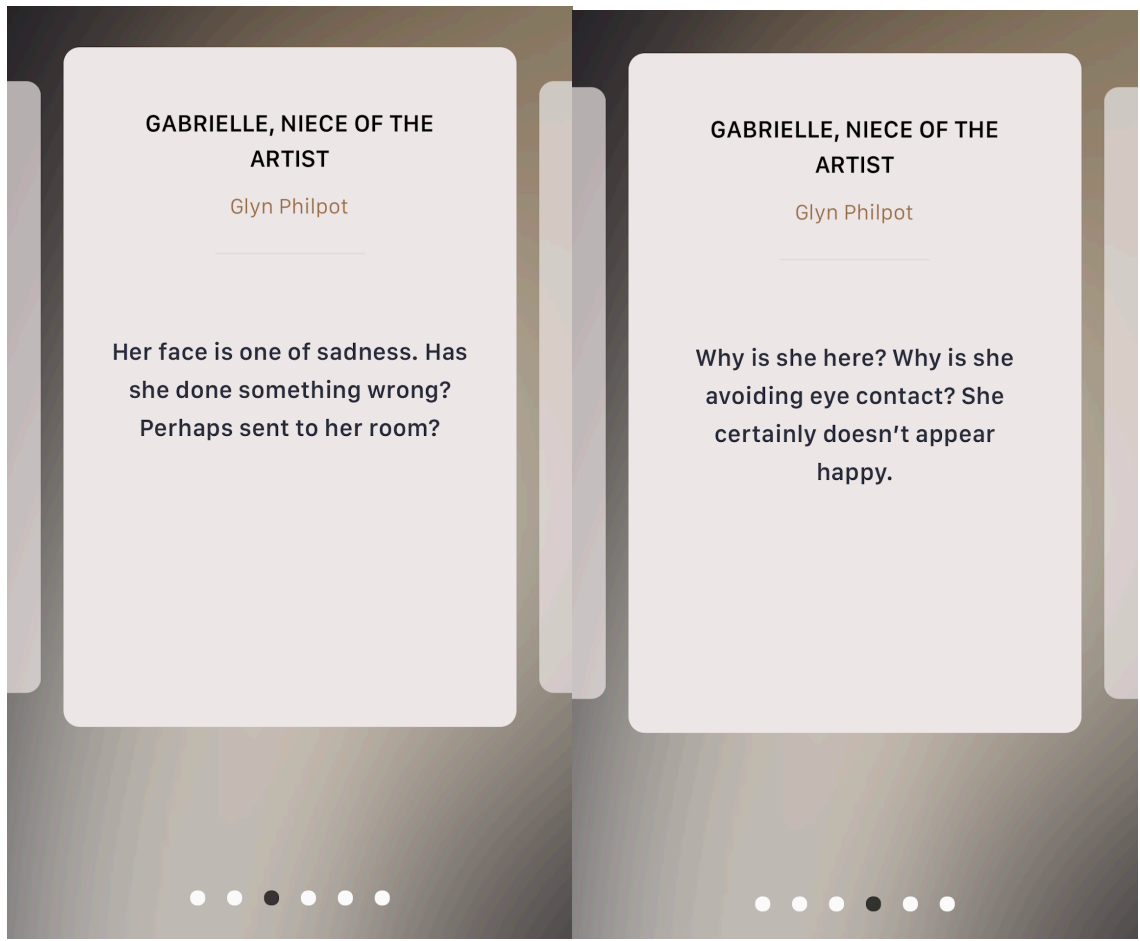

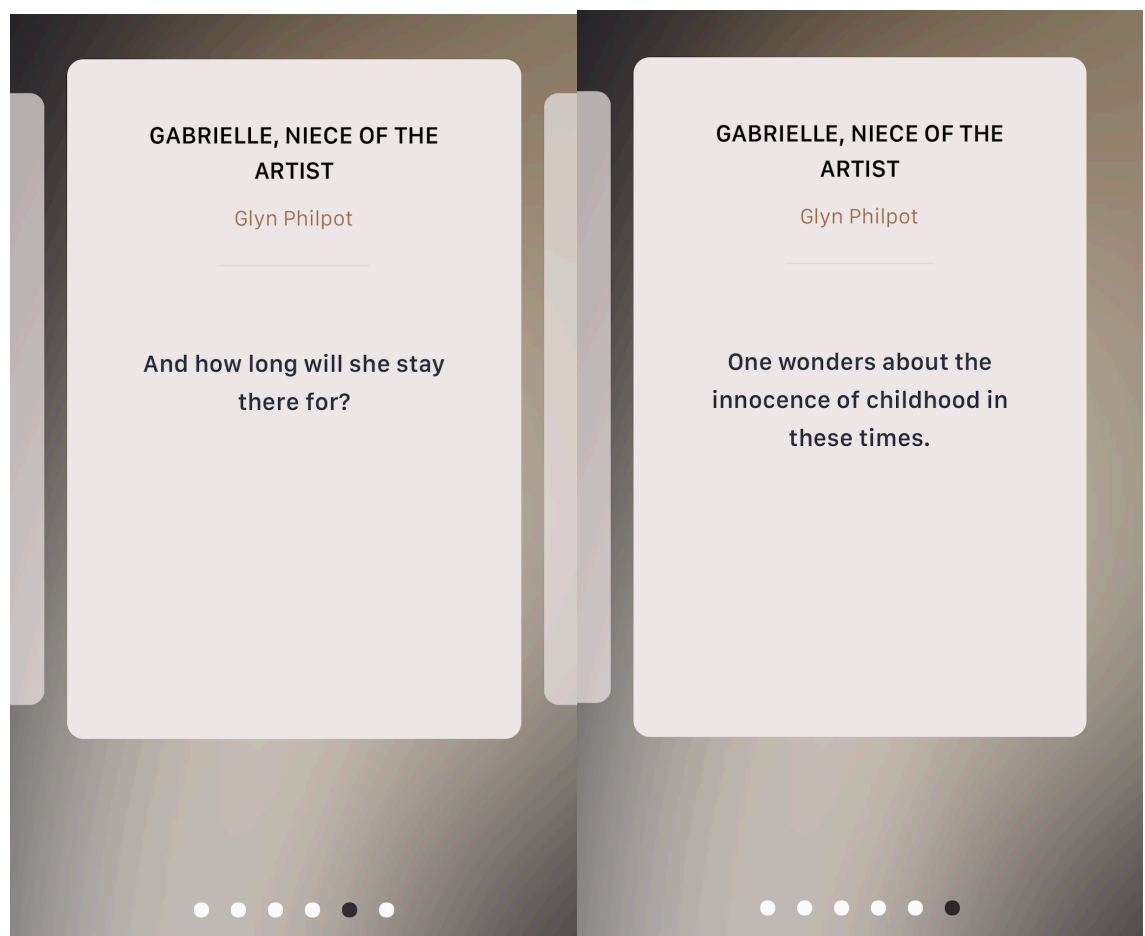

Figure K.4a - f: Text in the *One Minute Experience* app, about the painting Gabrielle, Niece of the Artist by Glyn Philpot. Royal Pavilion & Museums, Brighton & Hove.

According to Digital Manager Kevin Bacon, the experiment demonstrated the potential of the app to facilitate community co-production:

> "from past experience I have found that when invited to contribute to a museum tour or exhibition, people from outside of the museum often start writing and speaking as if they are a curator. That is a challenge if you are trying to move away from traditionally curatorial approaches to interpretation, or simply wish to capture an alternate tone of voice. One Minute Experience has the potential to address this problem, particularly if the use of the story editor can be built into a collaborative workshop."

Arguably, the *One Minute Experience* is a more museum-centric experience than *Never Let Me Go* and *Gift*. It is the museum that opens up for gathering stories and that also retains some editorial control over which stories will be available for other visitors. However, by guiding users away from adapting the traditional curatorial "voice", the app instead frees them up to be direct and personal, offering them an alibi to tell stories of their own liking, in their own voice, addressing other visitors.

A final, related example is the exhibition *Your Stories*, developed by the Serbian design company *NextGame* for the National Museum in Belgrade in order to explore how hybrid technologies could bring personal stories into the museum. They invited ordinary citizens to bring mundane objects to the museum to have them 3D scanned. The key instruction was to bring objects that were of great personal importance to them. The 3D scans were recreated as virtual objects, placed in close vicinity to physical museum objects that they, one way or another, could be considered related to. At each exhibit, visitors could scan a code to retrieve a 3D image of the object together with its background story.

Again it was the museum that collected and curated this exhibition, this time consisting of both objects and stories. However, an important difference was how it created a new relation between regular visitors and the museum: that of being a donor. Being able to add your personal possession to the museum collection was very appealing to visitors. The donors would talk fondly about this experience, and continued to stay updated and in touch throughout the process of exhibition design. For regular visitors who had not themselves contributed to the exhibition, the juxtaposition of modern objects and stories with the regular exhibits in the museum's ancient collection helped recontextualize the museum objects, provoking reflection on the value that those objects may once have held to their owners.

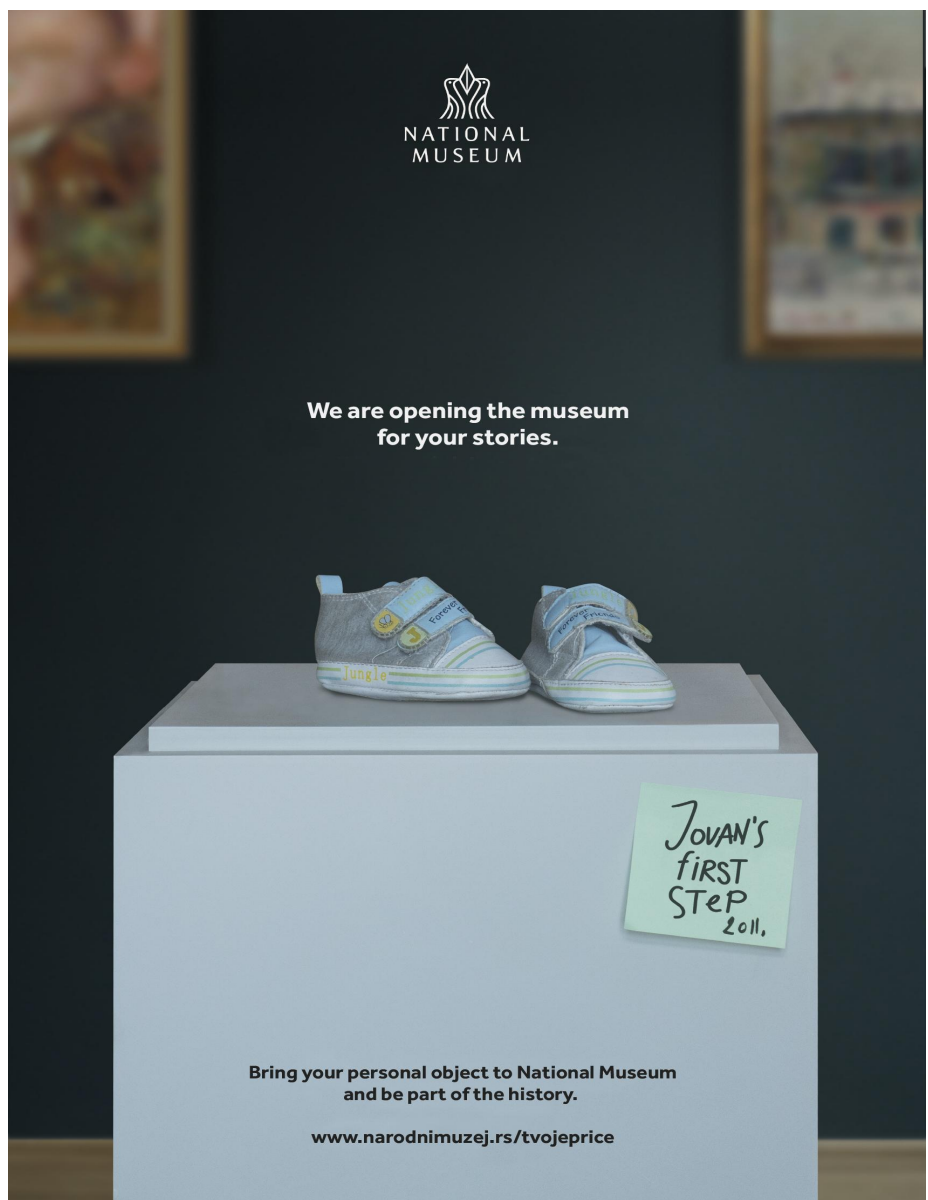

Figure K.5: Advertisement for the *Your Stories* exhibition by NextGame and the National Museum in Belgrade. Photo: NextGame Belgrade.

**Reflections**

The designs presented in this chapter all have in common that they foreground such meaning-making processes that take place between people, rather than between visitors and the museum. From the perspective of a museum curator or educator, this may provoke some concern: do these designs reduce the museum to merely a backdrop for socialising and play - an interactive stage for relationship work? Our answer is, first of all, that - like it or not - in reality most visitors already prioritise their social interactions. These designs offer ways in which interactions can be interpersonal, while also offering interesting encounters with museum exhibits. Second, some of these designs - *Never Let Me Go* and *Gift* - are also quite intimate, inviting people to share experiences on a one-to-one scale with someone special.

Thereby, they offer an alternative to the public (or semi-public) types of sharing that take place through social media - allowing for more profound and subtle ways of sharing an encounter with museum exhibits. As such, we believe that designing for interpersonal experiences holds the potential to facilitate visitor experiences that are not just engaging, but also deep and meaningful.

## Acknowledgments

The designs presented in this article are the result of extensive collaboration with more contributors than we can list here. We would like to particularly acknowledge the contributions from: Matt Adams, Kevin Bacon, Steve Benford, Dimitrios Darzentas, Ju Row Farr, Bogdan Spanjevic, Jocelyn Spence, Linda Stoltze, Nick Tandavanitj, Tim Wray. The GIFT project has received funding from the European Union's Horizon 2020 research and innovation programme under grant agreement No 727040.